\documentclass[prb,preprint,showpacs]{revtex4}
\usepackage{amsmath}
\usepackage{graphicx}

\begin{document}

\title{Valence modulations in CeRuSn}
\author{R. Feyerherm}
\email[Email of corresponding author: ]{ralf.feyerherm@helmholtz-berlin.de} 
\affiliation{Helmholtz-Zentrum Berlin f\"ur Materialien und Energie GmbH, BESSY, 12489 Berlin, Germany}
\author{E. Dudzik}
\affiliation{Helmholtz-Zentrum Berlin f\"ur Materialien und Energie GmbH, BESSY, 12489 Berlin, Germany}
\author{K. Proke{\v {s}}}
\affiliation{Helmholtz-Zentrum Berlin f\"ur Materialien und Energie GmbH,14109 Berlin, Germany}
\author{J. A. Mydosh}
\affiliation{Kamerlingh Onnes Laboratory, Leiden University, 2300RA Leiden, The Netherlands}
\author{Y.-K. Huang}
\affiliation{Van der Waals-Zeeman Institute, University of Amsterdam, Science park 904, 1098 XH Amsterdam, The Netherlands}
\author{R. P\"ottgen}
\affiliation{Institute for Inorganic and Analytical Chemistry, University of M\"unster, 48149 M\"unster, Germany}

\date{13.04.2014}

\begin{abstract}
CeRuSn exhibits an extraordinary room temperature structure at 300~K with coexistence of two types of Ce ions, namely trivalent Ce$^{3+}$ and intermediate valent Ce$^{(4-\delta)+}$, in a metallic environment. The ordered arrangement of these two Ce types on specific crystallographic sites results in a doubling of the unit cell along the $c$-axis with respect to the basic monoclinic CeCoAl-type structure. Below room temperature, structural modulation transitions with very broad hysteresis have been reported from measurements of various bulk properties. X-ray diffraction revealed that at low temperatures the doubling of the CeCoAl type structure is replaced by a different modulated ground state, approximating a near tripling of the basic CeCoAl cell. The transition is accompanied by a significant contraction of the $c$ axis. We present new x-ray absorption near-edge spectroscopy data at the Ce L$_{3}$ absorption edge, measured on a freshly cleaved surface of a CeRuSn single crystal. In contrast to a previous report, the new data exhibit small but significant variations as function of temperature that are consistent with a transition of a fraction of Ce$^{3+}$ ions to the intermediate valence state, analogous to the $\gamma \rightarrow \alpha$ transition in elemental cerium, when cooling through the structural transitions of CeRuSn. Such results in a valence-modulated state. 
\end{abstract}

 \maketitle

Cerium based intermetallic compounds are well known for their wealth of exotic electronic phenomena related to the instability of the sole 4$f$ electron of the Ce$^{3+}$ ion. For example, strong hybridization (with neighboring ions, $d$ bands) may lead to a dynamical mixture with the itinerant Ce$^{4+}$ state to a various extent, leading to an intermediate valence Ce$^{(4-\delta)+}$. A classical example is the $\gamma \rightarrow \alpha$ transition in elemental cerium, where a transition between a close to trivalent Ce state and an intermediate valence state can be driven by variation of temperature or application of pressure. This transition still is a topic of current research. \cite{Decremps, Lipp:2012} 

While most Ce-based intermetallics contain either trivalent or intermediate-valent Ce ions, for at least two compounds a 1:1 ordering of both species has been reported, namely CeRuSn\cite{Riecken:2007, Matar:2007, Mydosh:2011} and Ce$_2$RuZn$_4$.\cite{Mishra:2008, Eyert:2008}  The latter two materials are members of an interesting series of cerium-ruthenium based compounds that are characterized by a mixture of Ce ions with long and extraordinarily short Ce-Ru bonds on inequivalent Ce sites. The latter bonds can be as short as 2.23~\AA, much shorter than the sum of the covalent radii of Ce and Ru (2.89~\AA).\cite{Hermes:2009,Mishra:2011}

Among this extraordinary series, CeRuSn is unique by exhibiting temperature dependent structural modulations. At room temperature, two nonequivalent Ce sites, Ce1 and Ce2 exist, with the two shortest Ce-Ru bond lengths 2.33~\AA~ and 2.46~\AA~ for Ce1, and 2.88~\AA~ and 2.91~\AA~ for Ce2.\cite{Riecken:2007}  Correspondingly, the unit cell of CeRuSn at room temperature is doubled along the $c$ axis with respect to the parent CeCoAl-type structure (the so-called $2c$-superstructure). Band structure calculations \cite{Matar:2007} assign these two Ce sites to Ce$^{(4-\delta)+}$ (Ce1) and Ce$^{3+}$ (Ce2) species, which occupy 50\% of the total Ce sites each. Based on this picture, from x-ray absorption near-edge spectroscopy (XANES) data a valence of $\approx 3.36+$ has been derived for Ce1.\cite{Feyerherm}

This kind of charge-ordered superstructure is already surprising in view of the fact that CeRuSn has good metallic conductivity, and thus the Ce ions are embedded in a sea of conduction electrons. However, even more surprising is the observation of temperature dependent transitions to different structurally modulated  phases of CeRuSn, with quintupling below $\approx 290$~K and approximate tripling below $\approx 225$~K ($\approx$3$c$-superstructure) of the CeCoAl-type structure. Both transitions exhibit strong hysteresis and are associated with a significant volume collapse, mainly caused by contraction along the $c$ direction, $\Delta c/c \approx 0.8$\% between 340 and 180~K.\cite{Feyerherm, Praque:2012}     
While synchrotron x-ray and neutron diffraction data suggest that the ground-state structural modulation of CeRuSn is ill-defined\cite{Feyerherm,Prokes:PDF} or incommensurate and only close to a perfect tripling\cite{Prokes:3+1}, laboratory x-ray diffraction data have been interpreted in terms of an ideal tripling in the ground state.\cite{Praque:2012, Gribanova}  Whether this discrepancy is due to a sample quality issue remains unresolved.

Below $T_N = 2.7$~K, CeRuSn orders antiferromagnetically. While the original work\cite{Riecken:2007,Mydosh:2011} suggested that one half of the Ce ions order magnetically below $T_N $, in a  recent work,  based on a detailed low-temperature measurements of the magnetization and the specific heat, the scenario has been derived that below $T_N $ only one third of the Ce ions in CeRuSn are trivalent and carry a stable Ce$^{3+}$ magnetic moment, whereas the other two thirds of the Ce ions are in a nonmagnetic tetravalent and/or mixed valence state.\cite{Praque:2013} This picture is supported by the crystallographic data, which show that the idealized $3c$-superstructure is described by three nonequivalent Ce sites are present, with the two shortest Ce-Ru bonds 2.43~\AA~ and 2.43~\AA~ (Ce1), 2.92~\AA~ and 2.93~\AA~(Ce2), and 2.27~\AA~ and 2.75~\AA~ (Ce3).\cite{Praque:2012} The reduced bond lengths observed for Ce1 and Ce3 suggest that these two sites are occupied by Ce$^{(4-\delta)+}$. Together with the volume contraction, this behavior suggest that the structural transitions may be related to a valence transition of a fraction of Ce$^{3+}$ ions to the intermediate valence state, similar to the famous $\gamma \rightarrow \alpha$ transition in elemental cerium.  Notably, the occurrence of a valence transition in CeRuSn had already been suggested in Ref.~\cite{Riecken:2007} on the basis of $T$-dependent anomalies of magnetic susceptibility data. The magnetically ordered structure has been determined most recently by single crystal neutron diffraction. While the crystal structure was described by an incommensurate structural modulation with propagation vector $q = (0, 0, 0.35)$, the magnetic structure has exactly one-half that value, $q_{mag} = (0, 0, 0.175)$.\cite{Prokes:3+1}
Note that the structural modulation is very close to the commensurate wave vector (0, 0, 1/3), with a deviation of only 5\%.

In our previous report\cite{Feyerherm} we have presented XANES data that, in contrast to the above picture, revealed no significant variation of the average Ce valence with temperature. However, the growing evidence for the occurrence of a valence transition lead us to a re-examination of the problem. We noted that a possible source of error in the previous experiment was a minor quality of the sample surface. The XANES experiment was carried out at x-ray energies around $E = 5.72$~keV. At these energies, the penetration depth of photons is only $\approx \mu = 2 \mu$m, much smaller than the value of  $\mu = 17 \mu$m at $E = 12.4$~keV , which was used for the diffraction measurements in the same study. Therefore, any possible surface sample quality problems would have affected the XANES experiment more significantly than the diffraction data.

In the present Rapid Communication, we present new XANES data at the Ce L$_{3}$ absorption edge, measured on a freshly cleaved surface of a CeRuSn single crystal. In contrast to the previous data, the new XANES data exhibit small but significant variations as function of temperature that are consistent with a valence transition of a fraction of Ce$^{3+}$ ions to the intermediate valence state when cooling through the structural transitions.

The single crystal CeRuSn investigated in the present work is from the same batch (\#2) as that used our previous experiments.\cite{Feyerherm}  In contrast to the previous work, before each XANES experiment the surface of the sample was freshly cleaved at air (shiny face normal to the $c^{*}$ axis) and then immediately loaded into vacuum environment, so that the exposure to air was less than 5 minutes. Two independent Ce L$_{3}$ edge XANES experiments in total fluorescence yield mode have been carried out at the synchrotron source BESSY, one at the MAGS beamline\cite{MAGS}, and a second, independent cross-check experiment at  the instrument KMC-1. At MAGS, the sample was mounted to a displex cryostat that allows for experiments in the $T$ range $6 - 320$~K. The sample was first cooled to base temperature (6~K), in order to ensure that the sample is in the ground state, and then heated to 100~K and above before taking data. Special care has been taken that the sample was not exposed to x-rays at temperatures below 100~K, in order to avoid any x-ray induced effects.\cite{Feyerherm} The photon energy was calibrated with a Cr foil at the Cr K$_\alpha$ edge (5.989 keV). At KMC-1, the sample was mounted to a vacuum chamber, in which the sample was cooled by a liquid nitrogen system, which provides a base temperature of $\approx 100$~K. Both experiments yielded consistent results despite the fact that the cooling procedure at KMC-1 data was different. In the following we will only present data taken at MAGS.

\begin{figure} [t]
\includegraphics*  [width=4.5 in]  {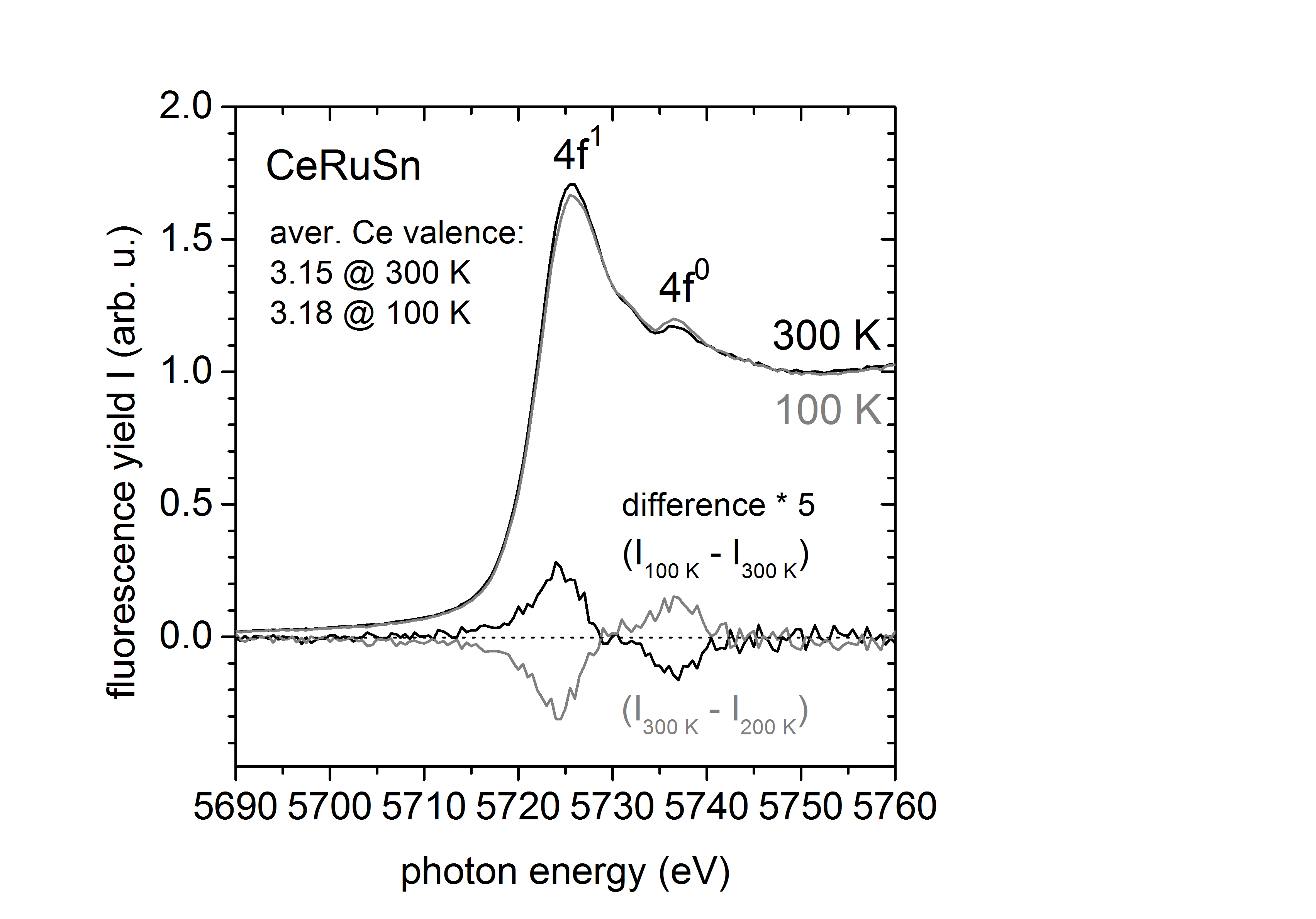}
\caption{Experimental XANES spectra at the Ce L$_3$ edge in CeRuSn at 100~K and 300~K. The ratio of intensities of the spectral features indicated by 4$f^1$ and 4$f^0$ allows an estimate of the Ce$^{3+}$/Ce$^{4+}$ ratio and thus of the average Ce valence. The difference between the 100~K and 300~K data is highlighted by the difference curve, where a second data set taken at 200~K is also shown. The $\approx 3\%$ reduction of the 4$f^1$ at the expense of the $4f^0$ feature observed at 100~K and 200~K is associated with a change of the average valence in the $\approx$3$c$ modulated ground state with respect to the $2c$ room temperature phase.}
\end{figure}

Figure~1 shows the fluorescence spectra measured across the Ce L$_3$ edge at 100~K and 320~K, i.e., in the $\approx$3$c$ modulated ground state and the $2c$ phase, respectively. Two main spectral features at at 5725.5 eV and 5735.5 eV are observed that are usually assigned to 4$f^1$ (Ce$^{3+}$) and $f^0$ (Ce$^{4+}$) states, respectively.\cite{Groot}
Clearly, a small but significant difference between the two spectra is observed, highlighted by the difference curve in Figure~1. (A similar difference curve determined from data taken at 200~K is also shown). The 4$f^1$ (Ce$^{3+}$) peak is reduced by $\approx 3\%$ at 100~K, while the 4$f^0$ (Ce$^{4+}$) peak is enhanced. The observed behavior has striking similarity with that observed at the $\gamma \rightarrow \alpha$ transition of Ce. There, an analogous shift of spectral weight has been reported when comparing the $\gamma$ (300~K) and $\alpha$ (20~K) phases of a Ce$_{0.93}$Sc$_{0.07}$ sample (the small Sc substitution stabilizes the $\alpha$ phase).\cite{Dallera}  In another experiment, Ce was driven through the $\gamma \rightarrow \alpha$ transition by applying a pressure of 20~kbar at room temperature, leading to quite significant changes of the XANES spectra.~\cite{Rueff:2006}

In x-ray absorption experiments an intermediate valence state, like Ce$^{(4-\delta)+}$, results in well separated spectral features, here 4$f^1$ and 4$f^0$, because the scattering event is much faster than the valence fluctuation time. The relative intensity of these spectral features can be directly extracted from the experimental data, taking into account a rounded two step-like background which simulates transitions to continuum states, where the relative heights of the two steps correlates with the spectral weights. Using this standard procedure, from the 320~K data we derive a ratio 0.83:0.17 (with error of 0.01) for the occupancy of 4$f^1$ and 4$f^0$ states. Naively, this would result in an average valence of 3.17(1)+, which is close to the value reported recently.\cite{Feyerherm}   However, small contributions from 4$f^2$ states, expected around 5718~eV and estimated to amount to 2-3\% of the total spectral weight for elemental Ce,\cite{Dallera, Rueff:2004}  cannot be resolved in the XANES experiment. Assuming that a similar contribution of 4$f^2$ states also exists in CeRuSn, we arrive at a more realistic estimate of the average valence of Ce in CeRuSn of 3.15(1)+ at 320~K. For the given structural model of the $2c$ phase (see above), assuming an ideal valence of 3+ for Ce2, this would lead to a Ce1 valence of 3.3+. We note, however, that another distribution of 4$f^0$ occupancy between Ce1 and Ce2 is possible (e.g., 3.25+ and 3.05+, respectively).

Accordingly, the $\approx 3\%$ reduction of the intensity of the 4$f^1$ feature at the expense of the 4$f^0$ can be directly translated into a change of the average valence from 3.15+ at 320~K to 3.18+ at 100~K. In the idealized crystallographic picture, the main difference between the $2c$ and the $\approx$3$c$ phases is the occurrence of a third crystallographic Ce site, Ce3, in the latter, with reduced Ce-Ru bond lengths. For Ce1 and Ce2, the bond lengths do not vary much and we may safely assume that their valence states do not change. Thus, we associate the observed variation of the average valence of CeRuSn between the $2c$ and $\approx$3$c$ phases with this new Ce3 site. Taking into account that the three Ce sites in the $\approx$3$c$ phase are equally populated, we obtain a valence of 3.24+ for Ce3. We note that this result implies that the structural transition is associated  by a valence transition from Ce$^{3+}$ to Ce$^{3.24+}$ of a fraction of only 1/6 of the Ce in CeRuSn. This interpretation remains generally valid even if the ground state is incommensurate with a periodicity close to $3c$. In that case, however, one has to consider that in the $\approx$3$c$ phase the Ce ions may exhibit a more complex distribution of valence states that \textit{on average} corresponds to 1/6 of the Ce having experienced the transition. In any case, the ground state exhibits a modulation of the Ce valence along the $c$ direction.

Considering that in CeRuSn only about 50\% of the volume is occupied by Ce ions (using covalent radii 165, 126, 141 pm for Ce,Ru,Sn, resp.) and that on average only 1/6 of the Ce exhibit a valence transition, the observed volume collapse of 0.8\% is quite considerable and would imply a local volume collapse around the Ce3 site by about 10\%. This has to be compared to the 14\% volume collapse in elemental Ce that, according to x-ray spectroscopy data,\cite{Dallera, Rueff:2006} is associated with a valence change from 3.03+ to 3.19+. Therefore, indeed the valence transition in 1/6 of the Ce in CeRuSn appears to be closely related  to the famous $\gamma \rightarrow \alpha$ transition in elemental Ce, a conjecture that has already been drawn.\cite{Praque:2012}

Thus, the physics of CeRuSn may be discussed in terms of the still active debate on elemental Ce, where presently the Kondo collapse scenario is prevailing.\cite{Lipp:2012}  In this scenario, in the collapsed state the moments (on Ce1, Ce3) are still mostly localized at the Ce sites but screened via the Kondo effect.

However, one also has to consider that already in the $2c$ phase (stable up to at least 800~K\cite{Feyerherm}) 50\% of the Ce ions are in a "collapsed" state and the temperature dependent valence transition constitutes only a gradual modification of the overall mixed Ce$^{3+}$ - Ce$^{(4-\delta)+}$ behavior of CeRuSn. Actually, the $2c$ phase may be regarded as a modulation of the valence state along the $c$ direction with propagation vector (0, 0, 0.5) with respect to the parent CeCoAl cell, where every second Ce is in an intermediate valence state. On cooling through the transition region, the fraction of collapsed Ce increases from 1/2 $\rightarrow$ 3/5 $\rightarrow$~$\approx$ 2/3 on cooling, leading to new valence modulations with associated propagation vectors (0, 0, 2/5) and (0, 0, $\approx$ 1/3), respectively. The valence modulation transitions are most probably driven by hybridization of the Ce $4f$ electron with the more extended Ru $4d$ electron wave functions. Apparently this hybridization varies significantly with temperature due to thermal expansion effects, resulting in the observed structural modulation transitions.

To summarize, our new XANES data on freshly cleaved surfaces of CeRuSn exhibits temperature dependent variations that are interpreted in terms of a valence transition of (on average) a fraction 1/6 of the Ce ions upon entering the modulated low-$T$ structure. 
This transition appears to be closely related  to the famous $\gamma \rightarrow \alpha$ transition in elemental Ce. However, CeRuSn exhibits unique behavior since in contrast to elemental Ce only a fraction of Ce ions in CeRuSn are in a "collapsed" state and this fraction varies with temperature, resulting in different valence-modulated phases. For a deeper understanding of this interesting behavior a systematic search for more Ce intermetallic compounds with similar phase transitions is desirable.

\textit{We thank M. Gorgoi for help with the KMC-1 experiment and W. Hermes (now at BASF) for part of the sample preparation.}

\end{document}